\documentclass[11pt]{article}

\usepackage[margin=0.95in]{geometry}
\usepackage{setspace}
\usepackage[T1]{fontenc}
\usepackage{newtxtext}
\usepackage{microtype}

\usepackage{amsmath,amssymb}

\usepackage{graphicx}
\usepackage{booktabs}
\usepackage{array}
\usepackage{multirow}
\usepackage{tabularx}

\usepackage{enumitem}
\setlist[itemize]{leftmargin=1.2em,itemsep=0.2em,topsep=0.3em}
\setlist[enumerate]{leftmargin=1.4em,itemsep=0.2em,topsep=0.3em}

\usepackage{titlesec}
\titleformat{\section}
  {\large\bfseries}
  {\thesection.}{0.5em}{}
\titleformat{\subsection}
  {\normalsize\bfseries}
  {\thesubsection}{0.5em}{}
\titlespacing*{\section}{0pt}{1.2ex plus 0.2ex minus 0.2ex}{0.6ex}
\titlespacing*{\subsection}{0pt}{0.9ex plus 0.2ex minus 0.2ex}{0.4ex}

\usepackage[font=small,labelfont=bf]{caption}
\usepackage{subcaption}

\usepackage[hidelinks]{hyperref}
\hypersetup{
  colorlinks=true,
  linkcolor=blue,
  citecolor=blue,
  urlcolor=blue
}

\usepackage{aas_macros}
\usepackage[numbers]{natbib}
\setcitestyle{aysep={},dashed=false}
\newcommand{\hst}{\textit{HST}}
\newcommand{\jwst}{\textit{JWST}}
\newcommand{\hwo}{\textit{HWO}}
\newcommand{\gaia}{\textit{Gaia}}
\renewcommand{\roman}{\textit{Roman}}

\renewcommand{\emph}[1]{\textcolor{magenta!80!black}{#1}}

\begin{document}

\begin{center}
{\bfseries\LARGE Hubble Astrometry for the Local Group and Beyond in the 2030s}\\[0.4em]
{\bfseries\large A White Paper for \textit{Building a Roadmap for Hubble Science into the 2030s}}
\end{center}

\vspace{-0.5em}
\hrule
\vspace{-0.3em}

\begin{center}
\begin{minipage}{0.95\textwidth}
\centering
\small
S. Tony Sohn\textsuperscript{1}, 
Paul Bennet\textsuperscript{1}, 
Kevin Andrew McKinnon\textsuperscript{2}, 
Roeland P. van der Marel\textsuperscript{1},
Mattia Libralato\textsuperscript{3},\\
Eduardo Vitral\textsuperscript{4},
Ekta Patel\textsuperscript{5}, 
Laura L. Watkins\textsuperscript{6}, 
Andr\'es del Pino\textsuperscript{7}, 
Andrea Bellini\textsuperscript{1}, 
Massimo Griggio\textsuperscript{1},\\
Mark A. Fardal\textsuperscript{8},
Nitya Kallivayalil\textsuperscript{9},
Jack T. Warfield\textsuperscript{9}, 
Karoline M. Gilbert\textsuperscript{1},  
Puragra Guhathakurta\textsuperscript{10}, \\
Daniel Weisz\textsuperscript{11},
Andrew Wetzel\textsuperscript{12}, 
Andrew B. Pace\textsuperscript{9}, 
Marcel S. Pawlowski\textsuperscript{13},
Joshua D. Simon\textsuperscript{14},\\
Gurtina Besla\textsuperscript{15},
Erik Tollerud\textsuperscript{1}, 
Xiaowei Ou\textsuperscript{9},
Niusha Ahvazi\textsuperscript{9},
Anna Bonaca\textsuperscript{14}
\\[0.75em]
\footnotesize
\textsuperscript{1}Space Telescope Science Institute (STScI), 
\textsuperscript{2}University of Toronto, 
\textsuperscript{3}INAF-Osservatorio Astronomico di Padova, \\
\textsuperscript{4}University of Edinburgh,
\textsuperscript{5}Villanova University,
\textsuperscript{6}AURA for the European Space Agency (ESA), STScI,\\ 
\textsuperscript{7}Instituto de Astrof\'isica de Andaluc\'ia (IAA-CSIC), 
\textsuperscript{8}Eureka Scientific,
\textsuperscript{9}University of Virginia, \\
\textsuperscript{10}UC Santa Cruz, 
\textsuperscript{11}UC Berkeley,
\textsuperscript{12}UC Davis, 
\textsuperscript{13}Leibniz-Institute for Astrophysics (IAP), \\
\textsuperscript{14}Observatories of the Carnegie Institution for Science, 
\textsuperscript{15}University of Arizona
\end{minipage}
\end{center}

\vspace{-1.5em}
\begin{center}
\textit{Submitted: May 22, 2026}
\vspace{-0.5em}
\end{center}

\begin{abstract}
\vspace{-0.5em}
\noindent Hubble’s long, stable astrometric baseline creates a rare opportunity for discovery in the Local Group and beyond. Many nearby galaxies, streams, and star clusters already have archival first-epoch imaging in hand, so future \hst\ observations over the next decade can turn those data into precise proper motions. For many Milky Way satellites, existing measurements already constrain orbital motion at a useful level, but \hst\ still offers a path to full 3D kinematics, internal motions, and more distant systems where current data remain insufficient. That opens the window to dynamical studies inaccessible through line-of-sight velocities alone, revealing orbital histories, internal kinematics, environmental processing, and the dark-matter structure of nearby galaxies. This white paper identifies \hst\ astrometry as an opportunity to capitalize on archival baselines by completing long-baseline measurements where first epochs already exist, establishing new first epochs where critical gaps remain, and assembling a legacy sample for future \jwst, \roman, and \hwo-era follow-up. The result will be a transformative dataset for the Local Group and Local Volume, driving discovery now while laying the groundwork for the next generation of dynamical studies for resolved stellar populations.\looseness=-2
\end{abstract}

\section{Motivation}
Most of our knowledge of kinematics has come from line-of-sight velocities, i.e., motions towards or away from us, which are more easily measured from ground-based observatories through spectroscopy. These measurements give us only one dimension of motion, thus severely limiting the characterization of kinematic properties. Proper motions (PMs) give us the transverse motions on the plane of the sky, providing the additional two missing components required for full 3D kinematics. The latter usually requires space-based observatories to circumvent uncertainties brought by atmospheric disturbance, and have generally been less precise than line-of-sight velocities. 
However, the two missing velocity components provide information that cannot be replaced simply by measuring lines-of-sight velocities more accurately \citep{Read2021}; they break key mathematical degeneracies \cite{Binney1982,vdMarel2010,Vitral2024} and provide answers to important questions that are impossible to address with only 1D motion \cite{Bennet2024,Vitral2026}. \emph{\hst's greatest remaining astrometric advantage is not only its spatial resolution and stability, but its unmatched ability to extend archival observations into the multi-decade baselines required for transformative PM science (Fig.\,1).}\looseness=-2

To obtain PMs, at least two epochs of observation are required separated by a time baseline. \hst\ remains exceptionally well suited to astrometric studies of resolved stellar populations in the Local Group, roughly the neighborhood out to $\sim$1 Mpc that includes the Milky Way, M31, and their satellite systems. For many galaxies, streams, and globular clusters, the key limitation is no longer whether first-epoch imaging exists, but whether those data are followed by new observations sufficiently separated in time to produce precise PMs. In many cases, the most valuable new observation is therefore not a deeper or higher resolution image but a new epoch that extends the baseline to two decades or more (see Fig.\,1).

That baseline matters because it determines the accuracy of the inferred PMs. \hst\ has already transformed Local Group science through deep imaging and star-formation histories, but many of the most important questions still require more accurate tangential motions. Adding that information turns a largely static picture of nearby systems into a fully dynamical view of Local Group evolution. Beyond the Local Group, neighboring groups in the Local Volume, roughly out to about 5--10 Mpc, offers the next frontier for these measurements and the chance to discover how environmental trends extend across a broader range of galaxies. This white paper identifies the Local Group and Local Volume PM subject area as a science case that requires \hst’s unique resolution, stability, and archival baselines over the next 10 to 15 years, while also providing anchor epochs and calibration foundations for future \hwo-era astrometry.

\section{\hst's Unique Astrometric Role}
\hst\ occupies a distinctive position in the astrometric landscape because it delivers the combination of stable imaging, fine spatial sampling, and well-characterized instrumentation with the depth required for precision work on faint resolved stars \cite{Rhodes2007,Ryon2024}. \gaia\ will continue to improve PMs for bright stars, but it is fundamentally limited in crowded fields and at faint magnitudes. These limitations largely restrict its reach to the Milky Way and its immediate surroundings \cite{Hodgkin_2021}, with some exceptions in nearby star-forming galaxies such as M31 and M33 \cite{vdMarel_2019}. In contrast, \hst\ can reach 5--10 magnitudes deeper over smaller fields while achieving comparable PM accuracies that enables extending measurements throughout the Local Group \cite{Sohn2020,Bennet2025,Patel2026} and into the Local Volume \cite{BennetHST2023,BennetHST2025}. This makes \hst\ the natural bridge between archival first epochs and the long-baseline measurements required for Local Group dynamics into the 2030s.

Controlling instrumental systematics becomes important as measurements approach the usual systematic floor, especially at large distances. The extensive \hst\ archive already provides many first epochs obtained with ACS and WFC3, and new observations in the same or closely matched configurations can minimize cross-instrument systematics and maximize the astrometric value of those baselines. In that sense, \emph{\hst\ astrometry is both a current science program and the temporal foundation for the next generation of dynamical astronomy.}\looseness=-2

\section{Core Science Drivers}
PMs provide the missing dynamical dimensions for Local Group astrophysics. They link resolved stellar populations to orbital histories \cite{Bennet2025}, turn dwarf galaxies into probes of halo structure and constrain the host’s mass \cite{Boylan-kolchin2013, Shen2022, Bhattacharya_2025}, and enable tests of galaxy evolution and dark matter physics that line-of-sight kinematics alone cannot provide \cite{Vitral2024,Vitral2026}. \hst\ is especially well positioned for this work because its archive already contains many of the first epochs needed for long-baseline astrometry (see Fig.~1). Since \hst’s astrometric characteristics are well calibrated and well understood \cite{Anderson2003,vanderMarel2007,Anderson2010}, new observations of these systems can immediately translate into transformative PM constraints. In many cases, the absence of a follow-up epoch therefore represents a missed opportunity, because the longest and most homogeneous PM baselines are still obtained with \hst\ itself.

\subsection{Dwarf-Galaxy Orbits and Local Group Assembly}
\hst\ PMs of dwarf galaxies provide one of the clearest routes to constraining their orbits and interaction histories with the Milky Way and M31 \citep[e.g.,][]{Sohn2020,Bennet2024, Bennet2025, Patel2026}. They distinguish first infall from long-term satellite membership, test whether apparent satellite associations are physically meaningful, and strengthen dynamical constraints on the mass profiles of both halos. With continuous growth of the satellite census, these measurements have become an increasingly important route to understanding assembly and environmental processing.\looseness=-2

\subsection{Internal Kinematics and Dark Matter in Dwarf Galaxies}
PMs of individual stars within dwarf galaxies provide information that line-of-sight data alone cannot recover. They help break degeneracies between mass profiles and orbital anisotropy \citep{Binney1982}, enabling robust measurements of the inner density profiles of some of the most dark-matter-dominated galaxies known\cite{Lokas2005,Massari2020,Yang2025}. One of the key outstanding questions in dark-matter physics is whether dark matter is self-interacting; if so, the inner density profiles of low-mass galaxies may be systematically different from predictions by collisionless models \citep{Sameie2020}. The next necessary step is to measure the full 3D internal kinematics of resolved stars. These measurements may provide the most promising observational route for probing self-interactions in the dark sector.\looseness=-2

\gaia\ can contribute only for the very closest dwarf galaxies, but \hst\ reaches deep enough to measure internal motions for a much larger and more distant sample \citep{Vitral2024,Vitral2026}. This makes \hst\ the key facility for building a statistically meaningful set of systems with full internal phase-space information, and therefore for connecting dark-matter physics to the internal structure of galaxies and the role of baryonic feedback.

\subsection{Orbits, Star-Formation Histories, and Environmental Processing}
\hst’s archive already contains deep color-magnitude diagrams and star-formation histories for many dwarf galaxies \cite[e.g.][]{Weisz_2011,Hidalgo2013,Albers2019}. PMs measured with \hst\ follow-up observations allow those stellar-population records to be interpreted in orbital context, linking quenching, gas loss, and structural change to infall times and pericentric passages \citep{Bennet2025,Patel2026}. This is especially valuable for ultra-faint and ancient systems, where star-formation histories alone cannot separate reionization-driven evolution from environmental effects \cite{Brown2014,Sacchi2021}.

\subsection{Stellar Streams and Halo Dynamics}
Stellar streams provide a complementary probe of the gravitational potential of the Milky Way and M31 \citep{Erkal2019,Shipp2021}. \hst\ PMs along streams constrain halo shape, mass distribution, and substructure, while also testing whether specific streams or substructures are physically associated \citep{Sohn2015}. PMs of individual stream stars improve membership selection, tighten stream models, and make it possible to trace the imprint of interactions with dark-matter subhalos \cite{Bonaca2019}. Because \hst\ reaches much fainter magnitudes than \gaia, it can extend stream studies to more distant and lower surface-brightness systems including streams near the outskirts of the Milky Way and M31 where the leverage on halo mass is especially strong. Although streams and dwarf galaxies probe different dynamical regimes, together they provide a powerful and complementary view of halo structure, substructure, and the dynamical history of the Local Group.

\subsection{Local Volume Galaxies}
While most current \hst\ astrometric programs focus on resolved stellar populations within the Local Group, the same long-baseline strategy can now be extended to nearby galaxy groups in the Local Volume. Beyond the Local Group, systems such as Cen A and M81 demonstrate that multi-decade \hst\ baselines are beginning to enable PM measurements at distances well beyond the Milky Way environment \citep{BennetHST2023,BennetHST2025}. Expanding these efforts in the coming decade would extend 3D kinematic studies to a much broader range of environments, allowing tests of whether trends in dwarf-galaxy evolution are universal across galaxy groups or specific to individual hosts. Such measurements would also make it possible to use the motions of entire galaxies to constrain the mass distribution in the Local Volume and compare it with cosmological models \cite{Libeskind2015,Wempe2026}. Unlike the Local Group, coverage of key Local Volume galaxies remains patchy (see Fig.~1). \hst\ observations would establish the first-epoch baselines required for future PM works, extending \hst’s astrometric reach into the 2030s and setting up the long-baseline measurements that future \hwo-era observations would complete.

\section{The Importance of the Next Several Cycles}

The scientific return from \hst\ PM work depends strongly on time baseline as well as image quality. PM uncertainties scale inversely with the baseline between observations, so extending a baseline from 10 to 20 years can improve PM precision by roughly a factor of two. For many Local Group systems, the most valuable observations over the next several cycles will be those that extend archival baselines as far as possible. A repeat observation can therefore have greater scientific value than a deeper single-epoch exposure. \emph{This opportunity is time sensitive}. \hst\ remains capable of delivering the stable, high-resolution imaging needed for precision astrometry, and continued operations into the early 2030s would preserve the most homogeneous baselines possible. Cross-observatory PM measurements have already been demonstrated to be feasible \citep{delPino2022,Libralato2023,Warfield2023,McKinnon2024}, and they will become increasingly important in the future, but they also introduce additional systematics. \emph{As long as \hst\ is available, the most accurate measurements for systems with archival \hst\ data will come from \hst-\hst\ baselines.} Observations obtained now will therefore define the longest and most internally consistent baselines that \hst\ itself can provide, and will serve as anchor epochs for later work with \jwst, \roman, and future facilities.\looseness=-2

\section{Program Concepts}

A coordinated \hst\ astrometry program can build a legacy PM network for resolved stellar systems across the Local Group. The core strategy is straightforward: obtain second-epoch imaging for targets with high-value archival first epochs, establish first epochs where critical gaps remain, and prioritize systems that maximize both immediate scientific return and long-term value for future missions. Highest priority should be given to systems with the longest archival baselines, strong ancillary datasets such as spectroscopy or star-formation histories, and science cases that remain inaccessible to \gaia. Additional priority should be placed on targets that provide strong synergies with future \jwst, \roman, and \hwo\ astrometric programs.

The highest priority is a set of archival second-epoch completion targets. These include nearby classical dwarf galaxies, selected ultra-faint dwarfs, and systems with especially strong ancillary data such as deep star-formation histories or existing spectroscopic constraints. For these targets, much of the investment has already been made, and a new epoch can transform archival imaging directly into precise dynamical measurements. The scientific return from these systems is immediate because the necessary first epochs already exist.\looseness=-2

A second component is first-epoch observations for scientifically important systems that still lack suitable ACS/WFC or WFC3/UVIS data for future PM work. This is especially timely because \textit{Rubin/LSST}, \roman, and \textit{Euclid} are expected to discover many new low-mass galaxies in the near future, greatly expanding the nearby discovery space. \hst\ can provide the necessary first-epoch astrometric baselines soon after discovery. This would allow these systems to be incorporated into long-baseline PM studies from the outset while maximizing their scientific return in the following decade. Unlike second-epoch completion targets, these observations represent strategic long-term investments that will mature alongside future flagship missions.

The program should also include special fields that provide strong leverage on halo dynamics, substructure, and cross-mission astrometric anchoring. Taken together, these components would define a community-scale astrometric survey that exploits \hst's archive, strengthens the scientific return of current observations, and establishes the baseline needed for \jwst-, \roman-, and \hwo-era dynamical studies.

\begin{figure}[!htbp]
\begin{center}
\includegraphics[width=0.93\linewidth]{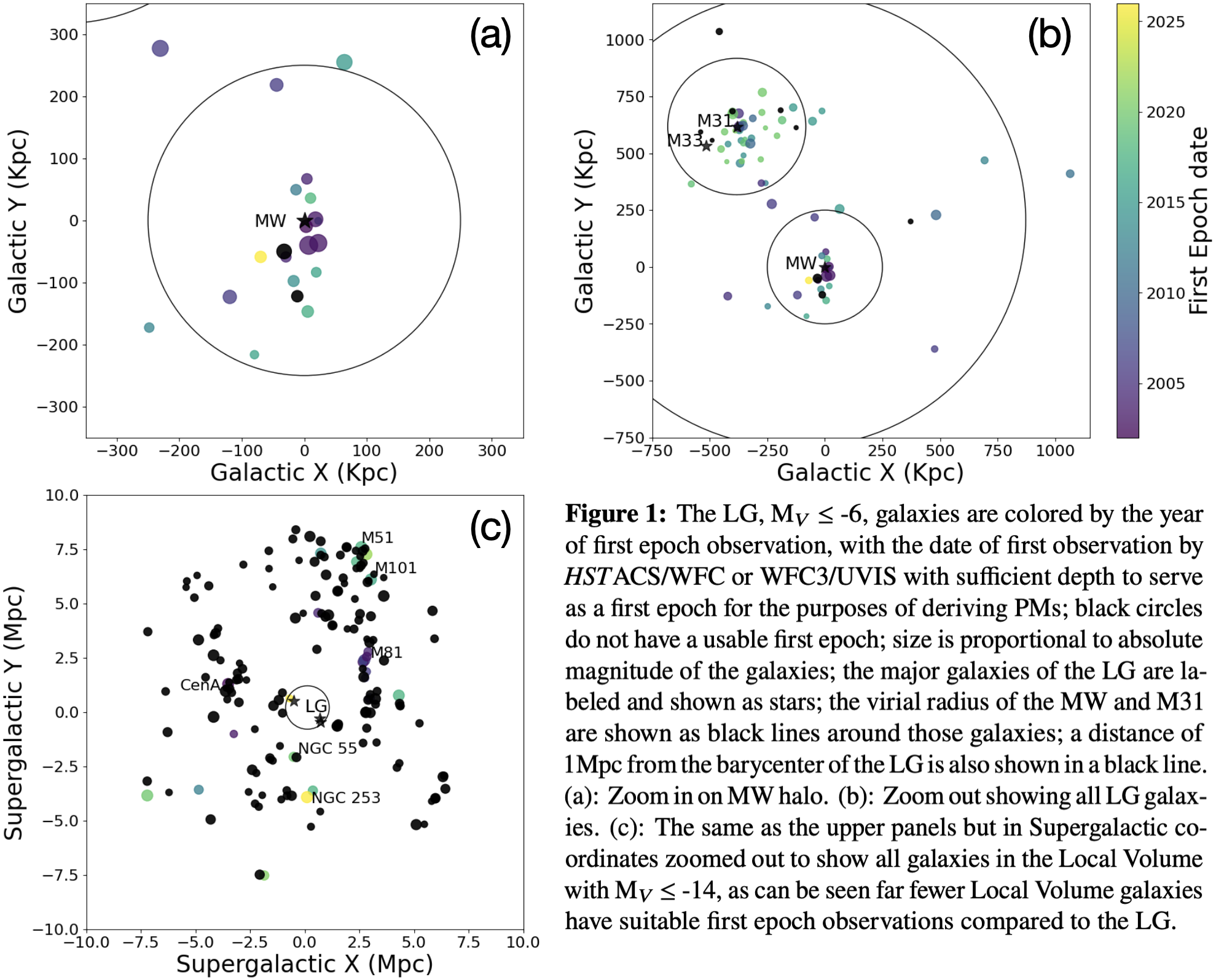} 
\label{Fig:Fig1}
\vspace{-2em}
\end{center}
\end{figure}

\section{Operational and Technical Requirements}

A successful astrometry initiative requires more than observing time. It also depends on an observing framework that recognizes the long-baseline structure of the science, preserves archival data, and supports the calibration stability needed for accurate PMs. The traditional GO framework is not well matched to PM science goals, which require observations separated by many years. \hst\ should therefore maintain a dedicated pathway for long-baseline and multi-cycle astrometric programs \citep{Jha2024}. Such mechanisms should support continuity across observing cycles and preserve the scientific value of archival first-epoch datasets. Furthermore, this observing framework must be agile enough to obtain immediate first-epoch imaging for new Local Group discoveries, ensuring we do not delay the start of their time baselines for future flagship missions.\looseness=-2

\emph{Continued operation of both ACS/WFC and WFC3/UVIS is crucial for preserving HST's long-baseline astrometric capability.} In particular, ACS/WFC is especially valuable because of its long operational history and the uniquely long time baselines it now enables, while WFC3/UVIS continues to provide complementary high-quality imaging for astrometric work. High-precision astrometry also depends critically on stable calibration and data infrastructure. Final astrometric precision depends heavily on distortion solutions, detector characterization, systematic-error control, and reproducible reduction methods. Moreover, charge-transfer efficiency (CTE) degradation in both ACS/WFC and WFC3/UVIS becomes increasingly important with instrument age, so continued calibration observations and updated correction strategies will be essential for preserving astrometric precision over the coming decade. Continued support for these elements is essential if HST is to remain a reliable anchor for multi-decade astrometry. A dedicated astrometric archive layer that not only identifies epochs, instruments, filters, and baseline lengths, but also flags the most scientifically valuable long-baseline opportunities and calibration requirements, would greatly expand the power of past and future \hst\ observations.\looseness=-2

\section{Synergies with \jwst, \roman, and \hwo}

\hst\ astrometry also plays a strategic role in preparing for future flagship missions. Its value lies not in duplicating the capabilities of \jwst, \roman, or \hwo, but in providing the early epochs and calibration anchors that later facilities will build on \citep{McKinnon2026}. \jwst\ is extending resolved-star astrometry into the infrared, \roman\ will add wide-field survey power, and future \hwo-era facilities may provide still greater reach, \emph{but none can recreate a long \hst\ baseline after the fact}.

Observations obtained over the next several cycles will therefore strengthen both current \hst\ science and the long-term value of later missions. They will increase the leverage of future \jwst\ and \roman\ measurements and provide anchor epochs for future flagship programs, making \hst\ astrometry a productive science program now while also laying the foundation for future astrometric studies.

\section{Concluding Remarks}

\hst\ archival imaging of resolved stellar populations can be evolved into a lasting astrometric legacy for the Local Group. Its resolution, stability, and multi-decade archive give it a time baseline no future facility can recreate once these epochs are gone. The available time baseline sets the ultimate limit on PM precision and therefore determines the scientific impact of these measurements. High precision PMs connect dwarf galaxies to their orbits, sharpen mass measurements of the Milky Way and M31, probe dark matter’s nature through dwarf galaxies and stellar streams, and place star-formation histories in their dynamical context. \hst\ observations taken now will also serve as anchor epochs for \jwst, \roman, and future flagship missions.

\smallskip
\smallskip
\noindent \emph{A coordinated \hst\ astrometry initiative would optimize urgency, reach, and long-term value, yielding both immediate science and a permanent legacy.}

\newpage
\setlength{\bibsep}{0.3em}
\bibliography{refs}

\end{document}